\documentclass[useAMS,usenatbib,usegraphicx,onecolumn]{mn2e}
\title[Gravitational lens mapping near fold caustic]{Asymptotic Expansions and  Amplification of a
Gravitational Lens Near a Fold Caustic}
\author[A.N.~Alexandrov, V.I.~Zhdanov]{A.N.~ Alexandrov$^{1}$ \thanks{E-mail:
alex@observ.univ.kiev.ua} and V.I.~Zhdanov$^{1,2}$
\thanks{E-mail: ValeryZhdanov@mail.ru}\\
$^{1}$ Astronomical Observatory, Taras Shevchenko National  University
of Kyiv, 3, Observatorna Str., Kyiv 04053,  Ukraine\\
$^{2}$ National Technical University of Ukraine ``Kyiv Polytechnic Institute", Kyiv 03056, Ukraine}

\def\y{\tilde y}
\def\x{\tilde x}
\def\*{\noindent}

\begin{document}

\date{Accepted 2011 June 21. Received 2011 May 25; in original form 2010 July 14}

\pagerange{541--554} 
\pubyear{2011}
\volume{417}

\maketitle

\label{firstpage}

\begin{abstract}
We propose two  methods that enable us to obtain
approximate solutions of the lens equation near a fold caustic
with an arbitrary degree of accuracy. We obtain ``post-linear''
corrections to the well-known formula in the linear caustic approximation
for the total amplification of two critical images of a point
source. In this case, in order to obtain the nontrivial
corrections we had to go beyond the approximation orders earlier
 used by Keeton et al. and to take into account the Taylor
expansion of the lens equation near caustic up to the fourth
order. Corresponding analytical expressions are derived for the amplification in cases of the Gaussian and power-law extended source models;
 the amplifications depend on three additional fitting
parameters. Conditions of neglecting
the correction terms are analysed. The modified formula for the  amplification
is applied to the fitting of light curves of the
 Q2237+0305 gravitational lens system in a vicinity of the high amplification events (HAEs).
 We show that the introduction of some ``post-linear'' corrections reduces $\chi^2$ by 30 per cent
 in the case of known HAE on the light curve
 of image C (1999). These corrections can be important for a precise comparison
 of different source models with regard for observational data.
\end{abstract}


\section{Introduction}
\noindent An extragalactic gravitational lens system (GLS) forms several images of a single quasar. The light from the quasar intersects a lensing galaxy in different
regions which correspond to different images. Variations of
gravitational fields in these regions due to stellar motions are
practically independent and lead to independent brightness
variations in different images (gravitational microlensing).
Comparison of the light curves of different images  allows one to
obtain a valuable information about the lens itself and about
the source as well \citep*{schneider_92,Wambsganss_06}. One of the important applications of this effect deals with a unique possibility to
study a fine structure of the central quasar region with the use of GLS. This idea first proposed by \cite* {grieger_88} appeals to the high amplification events (HAEs)
 in one of the images of
the quasar in GLS. An interesting  applications of HAEs
are known also in Galactic microlensing (see, e.g.
\citealt{Wambsganss_06}), in particular, using the caustic
crossing events for resolving of stellar profiles
\citep*{Bogdanov_02,dominik}. 

The conventional explanation of HAE relates it to the
caustic field in the source plane formed due to the inhomogeneous
gravitational field of a lensing galaxy on the line of sight of
the image \citep{schneider_92}. The source crossing of a caustic
leads to a considerable  enhancement of the image brightness, the crossing of a fold caustic being the most probable. The corresponding  variations of the brightness in a neighborhood of HAE can be approximately  described  by a formula containing a few
fitting parameters. This makes it possible to estimate  certain
 HAE characteristics, in particular, such as the source size
\citep{grieger_88}. For example, in the case of the well-known
Q2237+0305 GLS (Einstein Cross), several  HAEs was observed
\citep{wozniak_00, alcald_02, udalski_2006}, and the
estimates of the source size have been obtained within different
source  models \citep{wyithe_99,wyithe_00b,wyithe_00a,yonehara_01,shalyapin_01,shalyapin_02,Bogdanov_02}. Almost all HAEs in the  Q2237+0305 GLS are attributed just to a fold caustic
crossing in the source plane (see, e.g., \citealt{gil_06}). A possibility
to distinguish different source models is also discussed (
e.g. \citealt{goico_01}).

The lens equation near a fold can be expanded in powers of local
coordinates; in the lowest orders of this expansion, the caustic is
represented by a straight line; so, this approximation is often
referred as a ``linear caustic approximation''. In this
approximation, the point source flux amplification is given by a
simple formula, which depends on the distance to the caustic and
contains two parameters (e.g. \citealt{schneider_92}, \citealt{Cassan}). In most
cases, the linear caustic approximation is sufficient
to fit the observed light curves over the range of HAEs at the modern accuracy
of flux measurements.  The need for a
modification of this formula -- e.g., by taking  the
caustic curvature into account -- is, nevertheless, being discussed for a long time
\citep{Fluke,shalyapin_01,Pejcha}. We
hope that the future improvement of the photometric accuracy will make
it possible to obtain additional parameters of the lens mapping,
which are connected with the mass distribution in the lensing galaxy.
At the same time, we will show that the
 consideration of ``post-linear'' terms is sometimes
 appropriate to explain the present observational data.
We note that the corrections to the amplification factor in the case of macrolensing
 were the subject of investigations dealing with the problem
 of  ``anomalous flux ratios''  \citep*{keeton_05}.

Since the work by \cite{Kochanek_04} followed by a
number of works \citep*{Mortonson_05,gil_06,vakulik_07,anguita_08, Poindexter_08, Poindexter_10a, Poindexter_10b}, numerous statistical methods have been developed
to process complete light curves. This approach is very attractive because it allows one to consider
the whole aggregate of observational data on image
brightness variations in order to estimate the microlens masses
and the source model parameters. However, this treatment involves a
large number of realizations of the microlensing field, which  requires
a considerable computer time. On the other hand, the source
structure manifests itself only in HAEs; far from the caustic,
the source looks like a point one, and all the information about
its structure is lost. If we restrict ourselves to the HAE
neighborhood, then we use the most general model concerning a
microlensing field described by a small collection of Taylor coefficients in the
lens mapping. Therefore, the low-time-consuming semi-analytical
investigations dealing with caustic neighborhoods still preserve
their importance, not speaking about their use in computer codes.

In this paper, we propose relations for the total magnification of two
critical images of a point source in the first nontrivial
``post-linear'' approximations and use them to modify the
magnification of an extended source near the fold caustic.
In order to obtain non-zero corrections to the total
amplification of two critical images, we have to
consider additional higher-order terms in the expansion of
the lens mapping in comparison with the earlier works (see, e.g.,
\citealt{keeton_05}). Though the corrections are expected to be
small, they appear to be noticeable in some cases even in an
analysis of the existing data on light curves in
the Q2237+0305 GLS. The structure of this paper is as follows: after the
derivation of approximate solutions of the lens equation in
the required order, we deduce a formula for the magnification of a point source.
The result is used to obtain the magnification of a small
Gaussian source near the fold caustic.   We use the formula for the amplification of an extended source to fit the light curves
in GLS Q2237+0305. The obtained  post-linear corrections appear to
improve the agreement with observational data.

Appendices~\ref{A} and \ref{B} contain details on the approximation methods involved. Power-law source models are considered in Appendix~\ref{C}. Explicit comparison of our approximate formulas with 
counterparts from the paper by \citet{keeton_05} is made in Appendix~\ref{D}.

\section{Initial equations}
\label{s1}
First, we recall some general notions concerning
the gravitational lensing that can be found, e.g., in the book
by \cite{schneider_92}. The normalized lens equation has the form:

\begin{equation}
\label{eq1}
{\rm {\bmath y}} = {\rm {\bmath x}} - \nabla \Phi
\left( {\rm {\bmath x}} \right),
\end{equation}

\noindent where $\Phi \left( {\rm {\bmath x}} \right)$  is the lens
potential. This equation relates every point ${\rm {\bmath
x}}=(x_1,x_2)$ of the image plane to the point ${\rm {\bmath
y}}=(y_1,y_2)$  of the source plane. In the general case, there are
several solutions ${\bmath X}_{\left( i \right)} \left( {\bmath y}
\right)$ of the lens equation (\ref{eq1}) that represent images of
one point source at $\bmath y$; we denote the solution number by the index in parentheses.

If there is no continuous matter on the line of sight, the potential
must be a harmonic function $\Delta \Phi = 0$. Below, we will assume that this
condition is fulfilled in a neighborhood of the critical point.
We note, however, that if the
continuous matter density is supposedly constant during HAE, this can be taken into account by a suitable rescaling of the variables.

The amplification of a separate image of a point source is

\begin{equation}
\label{eq2}
K_{(i)} \left( {\rm {\bmath y}} \right) = 1
\mathord{\left/ {\vphantom {1 {\left| {J\left( {{\rm {\bmath
X}}_{\left( i \right)} \left( {\rm {\bmath y}} \right)} \right)}
\right|}}} \right. \kern-\nulldelimiterspace} {\left| {J\left(
{{\rm {\bmath X}}_{\left( i \right)} \left( {\rm {\bmath y}} \right)}
\right)} \right|},
\end{equation}

\noindent where $J\left( {\rm {\bmath x}} \right) \equiv \left|
{{D\left( {\rm {\bmath y}} \right)} \mathord{\left/ {\vphantom
{{D\left( {\rm {\bmath y}} \right)} {D\left( {\rm {\bmath x}} \right)}}}
\right. \kern-\nulldelimiterspace} {D\left( {\rm {\bmath x}}
\right)}} \right|$ is the Jacobian of the lens mapping. In the microlensing processes,
microimages cannot be observed separately; therefore, we
need the total amplification which is a sum of the amplification
coefficients of all the images.

The critical curves of the lens mapping~(\ref{eq1}) are determined by the
equation $J\left( {\rm {\bmath x}} \right) = 0$ and are mapped onto the
caustics in the source plane. The stable critical points of a two-dimensional mapping can be folds and cusps only, the folds being
more probable in HAE. In this paper, we confine ourselves to the
consideration of fold caustics. When a point source approaches the
fold caustic from its convex side, two of its images approach the
critical curve, and their magnification tends to infinity. They disappear
when the source crosses the caustic. These two images are called critical.

The standard consideration of the caustic crossing events deals
with the Taylor expansion of the potential near some point
$p_{cr}$ of the critical curve in the image plane. Let this point be the coordinate origin. We suppose that (\ref{eq1})
maps  $p_{cr}$ onto the coordinate origin of the source plane.
Further, we rotate synchronously the coordinate systems  until
the abscissa axis on the source plane be tangent to the caustic at the origin. The quantity $|y_2|$
defines locally  the distance to the caustic, and $y_1$ defines a
displacement along the tangent. For the harmonic potential, we can
write

\[
 y_1 = 2x_1 + a\left( {x_1 ^2 - x_2 ^2} \right) + 2bx_1 x_2 +
  c\left( {x_1 ^3 - 3x_1 x_2 ^2} \right) - d\left( {x_2 ^3 - 3x_2
x_1 ^2} \right) + gx_2 ^4 + ... ,
\]

\begin {equation}\label{eq3}
y_2 = \mbox{ } b\left( {x_1 ^2 - x_2 ^2} \right) - 2ax_1 x_2 +
  d\left( {x_1 ^3 - 3x_1 x_2 ^2} \right) + c\left( {x_2 ^3 - 3x_2
x_1 ^2} \right) + fx_2 ^4 + ...,
\end {equation}

\noindent where $a\mbox{, }b\mbox{, }c\mbox{, }d\mbox{, }g\mbox{,
}$ and $f$ are expansion coefficients. If  the $y_2$ axis is directed
toward the convexity of the caustic, then $b<0$
(at fold points, $b\neq0$).

\section{Expansion of the critical solutions in powers of a small parameter}
\label{s2}
\subsection{Method 1}
\label{ss2.1}
We now proceed to the derivation of approximate solutions of
Eqs.~(\ref{eq3}). To do this, we present  two different methods
which will be used to
have a possibility of mutual checks of cumbersome calculations.
The first method deals with analytical expansions in powers of a small
parameter. However, it results in nonanalytic functions of
coordinates leading to nonintegrable terms in the amplification
factor. The second method does not lead to such problems,
though it uses a somewhat more complicated representation of the
solution of the lens equation (containing square roots of
analytic functions). The methods agree with each other in a
common domain of validity; moreover, we use the second method to
justify some expressions in the amplification formulas
 in terms of distributions to validate applications to
 extended source models.

First, we use a regular procedure proposed by \cite*{alzh_03}
to construct solutions of Eqs.~(\ref{eq3}) with a desired
accuracy. This procedure is useful to study the light curve
of the point source which has a trajectory crossing the fold
caustic at some nonzero angle. We suppose that the source and
the caustic lie on different sides from the $y_1$ axis. Then, for $y_2
> 0$, we substitute

\begin {equation}
\label{eq4}
 y_i = t^2\y_i , \quad x_1 =
t^2\x_1 , \quad x_2 = t\x_2 ,
\end {equation}

\noindent where $i=1,2$, and $t$ can be considered as a parameter
of vicinity to the caustic. This is a formal substitution that
makes easier operations with different orders of the expansion.
After performing calculations, we shall put $t=1$ and thus return
to the initial variables $y_i$. However, if we put $\y_i$ to be
constant with varying $t$, then this substitution allows us to study
a local behaviour of critical image trajectories; $t=0$ corresponds
to  crossing the caustic by a point source,  and $t^2$ can be considered
as the time counted from the moment, when two critical images appear.
 Indeed, one can show \citep{alzh_03}
that two critical solutions ${\bmath X}_{i}(t)=(X_{\left( i
\right)1} ,X_{\left( i \right)2}) $ of the lens equation can be
represented by analytic functions of $t$ that, in the above
special coordinate system, have the behaviour $X_{\left( i
\right)1} \propto {t^2}$, $X_{\left( i \right)2}\propto
 t $ (see Appendix~\ref{A}). This allows us to look for  solutions of Eqs.~(\ref{eq3}), by using the expansions of $\x_i$ in
powers of $t$:

\[
 \x_1 = \x_{10} + \x_{11} t + \x_{12}\, t^2+...,
\]

\begin {equation}
\label{eq5}
 \x_2 = \x_{20} + \x_{21} t + \x_{22}\, t^2+....
\end {equation}

\* It should be stressed that the analyticity in $t$ does not mean
that the coefficients of expansions~(\ref{eq5}) will be analytic
functions of coordinates $\y_i$ in the source plane (see below).

In terms of the new variables~(\ref{eq4}), system~(\ref{eq3}) takes
the form (up to the terms $\sim t^2$)
\[
 \y_1 = 2 \x_1 - a \x_2^2 + t( {\,2b\x_1 \x_2 - d\x_2^3 } )
 + t^2\left( {a\x_1^2 - 3c\x_1 \x_2^2 + g\x_2^4 } \right),
\]

\begin {equation}
\label{eq25}
  \y_2 = - b\x_2^2 +  t( - 2a\x_1 \x_2 + c\x_2^3 )
 + t^2 ( b\x_1^2 - 3d\x_1 \x_2^2 + f\x_2^4 ).
\end {equation}

The substitution of expansions~(\ref{eq5}) into~(\ref{eq25}) allows us to determine all  coefficients successively. The results of
calculations are as follows.

For the zero-order terms:

\begin {equation}
\label{eq6}
\x_{10} =
\frac{1}{2}\left( {\y_1 - \frac{a}{b}\y{ }_2} \right), \quad
\x_{20} = \varepsilon \sqrt { {\y_2 } \mathord{\left/ {\vphantom
{{\y_2 } b}} \right. \kern-\nulldelimiterspace}
\left|b\,\right|}\;, \end {equation}

\noindent where $\varepsilon =\pm 1 $ determines two different
solutions.

The first-order terms are

\begin {equation} \label{eq5x} \x_{11} = - \varepsilon \sqrt
{\frac{\y_2}{\left|b\,\right|}} \frac{\left( {ac - aR^2 + bd}
\right)\y_2 + bR^2\y_1 }{2b^2}, \quad \quad \x_{21}
=\frac{1}{2}\left( {\frac{a^2 - c}{b^2}\y_2 - \frac{a}{b}\y{ }_1}
\right), \end {equation}

\noindent where $R^2 = a^2 + b^2$. The solutions up to this
accuracy has been obtained earlier by \cite{alzh_03} and
\cite{keeton_05} (see also \cite*{Congdon}). The contributions of this order are cancelled in
calculations of the total amplification factor of two
critical images. Therefore, to obtain a nontrivial correction to the
zero-order amplification, higher order approximations should be involved.

The second-order terms contain an expression nonanalytical in $\y_2$:

 \begin {equation}
\label{eq8}
\begin{array}{l}
 \x_{12} = {\displaystyle \frac{1}{4b^4}}\left( {3a^5 + 5a^3b^2 + 2ab^4 - 2b^3d - 2b^2g}
\right.
  + \vphantom{\displaystyle \frac{1}{1}}\left.{3ac^2 - 6a^3c + 3bcd - 8a^2bd + 2abf} \right)\y_2^2 + \smallskip\\
 \quad\quad\quad \quad\quad\quad\quad\quad\quad\quad\quad\quad\quad\quad\quad\quad\quad
  + {\displaystyle \frac{1}{2b^3}}\left( {2a^2c - b^2c + 3abd - 2a^2R^2 - } \right.
 \left. {b^2R^2} \right)\y_1\y_2 + {\displaystyle \frac{aR^2}{4b^2}\y_1^2} , \\
\end{array}
\end {equation}

\begin {equation}
\label{eq9}
\begin{array}{l}
 \x_{22} = \varepsilon \sqrt {{\displaystyle \frac{\y_2 }{\left| b \right|}}} \left[
{{\displaystyle \frac{1}{8b^3}}\left( {10a^2c - 5c^2 - } \right.}
\right. \vphantom{\displaystyle \frac{1}{1}}\left. {5a^2R^2 +
10abd - 4bf}\vphantom{R^2} \right)\y{ }_2 \left. { -
{\displaystyle \frac{3}{4b^2}}\left( {ac + bd - aR^2} \right)\y{
}_1 - {\displaystyle \frac{R^2}{8b}}{\displaystyle \frac{\y_{1 }^2
}{\y_2 }}} \right].
 \end{array}
 \end {equation}

\subsection{Method 2}
\label{ss2.2}

The second approach to the construction of approximate solutions of the lens equation in a vicinity of the fold is described in Appendix~\ref{B}. This allows us to provide the critical solutions of system~(\ref{eq25}) in the following form:

\begin {equation} \label{new1} \x_1 = p + tr\varepsilon \sqrt w , \quad \x_2 = t\bar s + \varepsilon \sqrt w ,\quad
\varepsilon = \pm 1. \end {equation}

  The iterative procedure described in  Appendix~\ref{B} yields analytical
expansions for the functions $p,r,\bar s,w $ both in powers of $t$ and
$\y_i$. The application of this method to system~(\ref{eq25})
gives (up to the terms $\sim t^2$)

\begin {equation}
\label{pq}
p = \x_{10} + t^2\x_{12} ,
\end {equation}

\* where $\x_{10}$ and $\x_{12}$ are given by relations~(\ref{eq6}) and (\ref{eq8}), and

\begin {equation} \label{sw}
r = -
\frac{R^2\y_1 }{2b} + \frac{\y{ }_2}{2b^2}\left[ {aR^2 - (ac + bd)}
\right], \quad \bar s = - \frac{a}{2b}\y_1 + \frac{a^2 - c}{2b^2}\y_2 , \end {equation}

\begin {equation} \label{new1Z} w = - \frac{\y_2}{b} + \frac{t^2}{4b^2}\left[R^2{\y_1}^2 + \frac{6}{b}\left( {
bd-ab^2+ac-a^3} \right)\y_1 \y_2 + \frac{1}{b^2}\left(
{5a^2(R^2 - 2c) + 5c^2 - 10abd + 4bf} \right){\y_2}^2\right]. \end {equation}

As is seen, all these expressions are analytic functions of both $t$
and $\y_i$. If we expand $\sqrt w$ in powers of $t$,
then we immediately have the solution in the form~(\ref{eq5}) with
coefficients~(\ref{eq6}), (\ref{eq5x}), (\ref{eq8}), and
(\ref{eq9}).

\section{Amplification factor} \label{s3}
\subsection{Point source}\label{ss3.1}

For the Jacobians $J\left( {{\rm {\bmath X}}_{\left( i \right)} \left(
{t^2{\rm {\bmath \y}}} \right)} \right)$ ($i = 1,2$ corresponds to
$\varepsilon = \pm 1)$, we obtain up to the terms $\sim t^3$:

 \[J = 4\varepsilon t\sqrt {\left|b\right|\y_2 } + 4t^2 {\displaystyle \frac{\left( {R^2 - c}
\right)}{b}}\y_2 \;
 + \varepsilon t^3\sqrt {\left|b\right|\y_2 } \left[ { - 3 {\displaystyle \frac{a\left( {R^2 -
c} \right) - bd}{b^2}}\y_1 } \right. +
\]
\begin {equation}
\label{eq10}
\quad\quad\quad\quad\quad\quad\quad\quad\quad
 \mbox{ } + {\displaystyle \frac{1}{2b^3}}\left(7a^2R^2 - 8cR^2 + 7c^2 - 6a^2c \right.
 \left. \vphantom{\displaystyle \frac{\left( {R^2}\right)}{b^2}}
 \left. -\:30abd + 24b^2c +
12bf\right) \y_2  - {\displaystyle \frac{R^2}{2b}\frac{\y_1^2
}{\y_2 }}\right].
\end {equation}

According to (\ref{eq2}), the value of $J^{ - 1}$ yields the
amplification of individual images. Note that the terms up to order
$\sim t^2 $ were obtained by \cite{keeton_05}. The final result
for the total amplification of two critical images (in terms of
the initial variables $y_i$ after putting $t=1$) is as follows:

\begin {equation} \label{point amplification} K_{cr} = \frac{1}{2}\frac{\Theta
\left( {y_2 } \right)}{\sqrt {\left| b \right|y_2 } }\left[ {1 +
Py_2 + Qy_1 - \frac{\kappa }{4}\frac{y_1^2 }{y_2 }\mbox{ }}
\right], \end {equation}

\noindent where
\begin {equation} \label{eq12}
 P = 2\kappa +  \frac{15}{8| b |^3}\left[ {a^2 b^2 + ( a^2 -
c)^2} \right] -  \frac{3}{4b^2}( {2f - 5ad}), \quad\quad
 Q = \frac{3}{4b^2}\left( {a^3 - ac + ab^2 - bd} \right),
\quad \quad \kappa = {\displaystyle \frac{a^2 + b^2}{2\left| b
\right|}} \, ,
\end {equation}

\*$\Theta
{(y_2) }$ is the Heaviside step function. Note that $\kappa$ is the
caustic curvature at the origin \citep {Gaudi,alzh_03}
which enters explicitly into the amplification formula.

Formula~(\ref{point amplification}) yields an effective
approximation for the point source magnification near the coordinate
origin provided that $y_2> 0$, and ${y_2}/{y_1^2}$ is not too small
(see the term containing $\kappa$). For a fixed source position, this
 can be satisfied always by an appropriate choice of the coordinate
origin, so that the source will be situated almost on a normal
to the tangent to the caustic.

If the source is on the caustic tangent or in the region
between the caustic and the tangent, then formula~(\ref{point
amplification}) does not represent a good approximation to the point
source
 magnification. Nevertheless, in case of an extended
 source, we will show that
result~(\ref{point amplification}) can be used to obtain
approximations to the amplification of this source even as it
intersects the caustic. However, to do this, we need to redefine
correctly the convolution of (\ref{point amplification}) with a
brightness distribution.

\subsection{Transition to extended source} \label{ss3.2}
Let  $I({\bmath y})$ be a surface brightness distribution of an
extended source.  If the source center is located at the point
${\bmath{Y}}=(Y_1,Y_2)$ in the source plane, then the total
microlensed flux from the source is
\begin {equation}
\label{flux_extended}
 F( {\bmath Y})= \int\!\!\!\int I ( { {\bmath y}( {\bmath x} )-{\bmath Y}})\, dx_1 dx_2
 = \int\!\!\!\int {K( {\rm {\bmath y}} )
 I ( {\rm {\bmath y}-{\bmath Y}})\,dy_1 dy_2},
\end {equation}
where the point source amplification $K( {\bmath y} ) =
\sum\limits_i {K_i } $ is the sum of amplifications of all the images. The result of using  the first integral from Eq.~(\ref{flux_extended}) obviously is equivalent to the result of the well-known ray-tracing method \citep{schneider_92} (when the pixel sizes tend to zero).
Near a caustic, one can approximate $K( {\bmath y} )=K_0 + K_{cr} (
{\bmath y})$, where $K_0 $ is an amplification of all noncritical
images that is supposed to be constant during HAE, and $K_{cr}$ is
the amplification of the critical images.

Formula~(\ref{point amplification}) contains the non-integrable
term $\sim\Theta(y_{2})  ( {y_2 })^{ - 3 /2} $. Therefore, the
question arises of how formula~(\ref{point amplification}) can be used
in situation when the extended source intersects a caustic and
some part of the source is in the zone between the tangent and the
caustic. In view of Section~\ref{ss2.2}, it is evident that the
mentioned term is a result of the expansion of the root $\sqrt
{y_2 + \kappa y_1^2 t^2/2+...}$  in the approximate solution
(\ref{new1}-\ref{new1Z}).  Any non-integrable terms in $K_{cr}$
does not arise without using this expansion. It
is easy to show that, in order to define $K_{cr}$ correctly, one
must replace  the term $\Theta(y_2)
( y_2 )^{ - 3 /2} $ in~(\ref{point amplification}) by the distribution  (generalized function)
$( y_2 )_+^{ - 3/2}$  \citep{Gel'fand_64}. We recall that the
distribution $ y _+^{ - 3/2}$ of the variable $y$ is defined by
the expression
\[
\int{ y _ + ^{-3 / 2} f( y)dy} = \int\limits_0^\infty { \frac{f( y
) - f( 0 )}{y ^{3/2} }} dy = 2\int\limits_0^\infty {y^{- 1/ 2}}
\frac{\partial f( y)}{\partial y}dy
\]

\* for any test function $f(y)$.

After this redefinition, we have

 \begin {equation}
\label{generalized Kcr} K_{cr} = {\displaystyle \frac{\Theta
\left( {y_2 } \right)}{2\sqrt {\left| b \right|y_2 } }}\left[ {1 +
Py_2 + Qy_1 \mbox{ }} \right] - {\displaystyle \frac{\kappa
}{8\sqrt {\left| b \right|} }}{y_1^2 }{\left( y_2 \right)_ + ^{-3
/ 2} }. \end {equation}

\*This formula can be  used to correctly derive an approximate
magnification of a sufficiently smooth extended source including
the case where the source crosses the caustic.

\subsection{Extended Gaussian source} \label{ss3.3}

Now we use formula~(\ref{generalized Kcr}) to derive the
magnification of a Gaussian source with the brightness
distribution

\begin {equation} \label{gaussian distribution}
 I_G ({\rm {\bmath y}}) =
{\displaystyle \frac{1}{\pi L^2}}\exp ( - {\bmath y}^2/L^2) ,
\end {equation}

\* where the parameter $L$ characterizes the source size.

The amplification of an extended  source is defined as the ratio
of the lensed flux~(\ref{flux_extended}) to the flux of the
unlensed source $F_0 = \int\!\!\!\int I \left( {\rm {\bmath y}}
\right)dy_1 dy_2 $ which is equal to 1 in case of
formula~(\ref{gaussian distribution}). The amplification of
a Gaussian source $K_G$ is obtained by the substitution of
(\ref{gaussian distribution}) and (\ref{generalized Kcr}) into
(\ref{flux_extended}).

Further, we introduce the dimensionless coordinates $s = Y_1/L,\, h
= Y_2/ L$ of the source centre and the functions

\begin{equation}
\label{eq16}
 I_k \left( h \right) = \int\limits_0^\infty {u^{k - 1 \mathord{\left/
{\vphantom {1 2}} \right. \kern-\nulldelimiterspace} 2}\exp \left(
{ - u^2 + 2uh} \right)du} =  {\displaystyle
\frac{1}{2}}\sum\limits_{n = 0}^\infty {{\displaystyle
\frac{\Gamma \left( {\textstyle{1 \over 4} + \textstyle{{k + n}
\over 2}} \right)}{n!}}} ( {2h})^n.
\end{equation}

These functions can be expressed in terms of the confluent
hypergeometric function  $_1F_1 $ or the parabolic cylinder function $D$ \citep{bateman}:

\begin{equation}
\label{eq17}
 I_k \left( h \right) = {\displaystyle\frac{1}{2}}\Gamma \left( {\textstyle{1 \over 4} +
\textstyle{k \over 2}} \right){ }_1F{ }_1\left( {\textstyle{1
\over 4} + \textstyle{k \over 2},\textstyle{1 \over 2};h^2}
\right)  + h\Gamma \left( {\textstyle{3 \over 4} + \textstyle{k
\over 2}} \right){ }_1F{ }_1\left( {\textstyle{3 \over 4} +
\textstyle{k \over 2},\textstyle{3 \over 2};h^2} \right)
 = 2^{ - \left( {\frac{k}{2} + \frac{1}{4}} \right)}\Gamma \left( {k +
\frac{1}{2}} \right)e^{\frac{h^2}{2}}D_{ - \left( {k +
\frac{1}{2}} \right)} \left( { - \sqrt 2 \cdot h} \right)\mbox{ .}
\end{equation}

The substitution of (\ref{generalized Kcr}) and (\ref{gaussian distribution}) in~(\ref{flux_extended}) yields
\begin{equation}
\label{eq18}
 K_G \left( {s,h} \right) = {\displaystyle \frac{1}{2\sqrt {\pi \left| b \right|L}
}}\left\{ \vphantom {\left [{\displaystyle \frac{\kappa
}{2}}\right]} \Phi_0 \left( h \right) + L\left[ \vphantom
{{\displaystyle \frac{\kappa }{2}}}P \Phi_1 \left( h \right) -
\right. \right.
 {\displaystyle \frac{\kappa }{2}} \Phi_2\left( h \right) + Q  s  \Phi_0 \left( h \right) \left.
{\left. {- \;\vphantom {{\displaystyle \frac{\kappa }{2}}}\kappa
s^2 \Phi_2}\left( h \right) \right]} \right\} \mbox{ }.
\end{equation}
\noindent
Here,
\begin {equation}
\label{Fi_i}
 \Phi _0 \left( h \right) = I_0 \left( h \right)\exp \left( { - h^2} \right),
\quad
\Phi _1 \left( h \right) = I_1 \left( h \right)\exp \left( { - h^2}
\right),\quad\Phi _2 \left( h \right) = \left[ {hI_0 \left( h \right) - I_1 \left( h
\right)} \right]\exp \left( { - h^2} \right).
 \end {equation}
\*
We have checked this result by the direct substitution of expansions~(\ref{eq25}) in the
first integral of Eq.~(\ref{flux_extended}), assuming $t = \sqrt L $ and expanding the resulting
integral in powers of this parameter up to the second order. Note that the main term of
(\ref{eq18}) which corresponds to the linear caustic approximation was first
obtained by \cite{Schneider Weiss}.

Analogous considerations allowed us to obtain formulas for the  magnification of extended sources for two types of power-law brightness profiles (Appendix~\ref{C}); the results are represented analytically in terms of hypergeometric functions.

\begin{figure}
    \centering
        \includegraphics{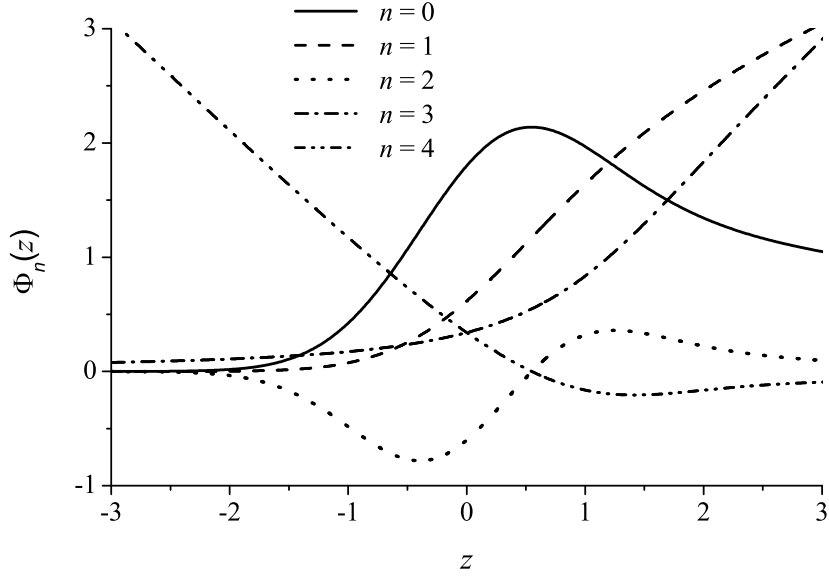}
    \caption{Behaviour of individual functions (\protect\ref{Fi_i}) which generate dependence (\protect\ref{eq18})
and the ratios (\protect\ref{Fi_34}) of correction functions  to the main one.}
    \label{Fi1}
\end{figure}

Let us discuss formula~(\ref{eq18}) in more details.
The functions $\Phi _i \left( h \right)$ are shown in Fig.~\ref{Fi1}.
The distinctive variations of these functions take place for $-2L<y_2<2L$.
For $y_2 < - 2L$, they are practically equal to zero, and, for $y_2 > 2L$, they
have the following asymptotic behavior:

\begin {equation}
\label{Asymp}
\Phi _0 \left( h \right) \cong {\displaystyle\sqrt {\frac{\pi }{h}}} ,
\quad
\Phi _1 \left( h \right) \cong {\displaystyle\sqrt {\pi h}} ,
\quad
\Phi _2 \left( h \right) \cong {\displaystyle\frac{1}{4h}\sqrt {\frac{\pi}{h}}} .
 \end {equation}

 We also introduce the functions
\begin {equation}
\label{Fi_34}
\Phi _3 \left( h \right) = {\Phi _1 } \mathord{\left/ {\vphantom {{\Phi _1 }
{\Phi _0 }}} \right. \kern-\nulldelimiterspace} {\Phi _0 },
\quad
\Phi _4 \left( h \right) = - {\Phi _2 } \mathord{\left/ {\vphantom {{\Phi _2
} {\Phi _0 }}} \right. \kern-\nulldelimiterspace} {\Phi _0 },
 \end {equation}
\*
which are designed to estimate the
contribution of the correction terms. These functions are also  shown in Fig.~\ref{Fi1}. At the origin, we have $\Phi _3 (0) =
\Phi _4 (0) = 0.338$.

A special attention must be given to the term which is proportional to
the monotonically increasing function $\Phi _1 \left( h \right)$. This
correction becomes especially noticeable on the inner side of the caustic,
its sign being determined by the sign of the parameter $P$ from Eqs.~(\ref{eq12}). Such behaviour
allows us to hope that the determination of this term from the observational
data will not be too difficult.

\begin{figure}
    \centering
        \includegraphics{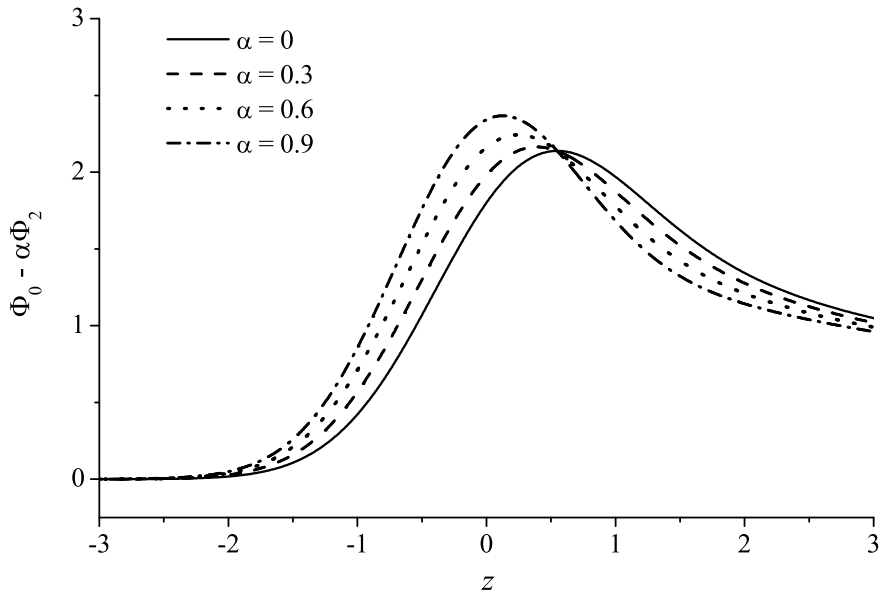}
    \caption{Influence of the correction term proportional to $\Phi _2 $.}
    \label{Fi2}
\end{figure}

The effect of the third term of formula~(\ref{eq18}) that is
proportional to $\Phi _2 \left( h \right)$ is shown in Fig.~2 by the dependence $\Phi _0 \left( h \right)~-~\alpha \Phi _2 \left( h \right)$ for
various values of the coefficient $\alpha$. According to (\ref{eq18}), the parameter $\alpha=L\kappa/2$ equals to half of the ratio of the source radius to the curvature radius
of the caustic at the origin. One can see from Fig.~\ref{Fi2} that this term can noticeably affect the
determination of the source size and the time moment of crossing the caustic .

The last two terms of Eq~(\ref{eq18}) contain $s\propto y{ }_1$ and
 $s^2\propto {y{ }_1}^2$. For a fixed position of the source, one can exclude
these terms by the origin displacement. In case of the rectilinear motion of a source,
the impact of these terms can be noticeable for small angles of the intersection of the source trajectory  with the caustic.

We now discuss the conditions of applicability of the approximation
methods involved for modelling the light curves in a vicinity of HAE.
For example, with regard for observational data, we can require that, in the interval $ - 0.5 \le h \le 1$, the
contribution of each correction do not exceed five per cent. We see from
Fig.~\ref{Fi1} that, in this interval, $\Phi _3 \left( h \right) < \Phi _3 \left( 1 \right)
\approx 0.83$ and $\left| {\Phi _4 \left( h \right)} \right| < \Phi _4
\left( { - 0.5} \right) \approx 0.73$. From the condition of smallness of
the corrections, we find the following restrictions on the parameters of the
model:

\begin {equation}
\label{LCA}
LP < 0.06,
\quad
L\kappa < 0.13,
\quad
LQ \cdot \cot \left( \beta \right) < 0.05,
\quad
L\kappa \cdot \cot ^2\left( \beta \right) < 0.27,
 \end {equation}

\noindent
where $\beta $  is the angle between the source trajectory and the caustic.
Under these conditions and within the specified margin of error,  one can
use the linear caustic approximation. Thus, in addition to the
requirement of the smallness of the caustic curvature, we obtain some extra restrictions on the possible variations of the lens potential on the scale
of the source size. It is clear that the strengthening of requirements for the model accuracy and/or the extension
of the interval of $h$ leads to the strengthening of the conditions found.

Clearly, formulas~(\ref{generalized Kcr}) and (\ref{eq18}) make sense only if we can ignore the discarded high-order terms. This means, in turn,  that the correction terms in (\ref{eq18}) themselves must be sufficiently
small.

\section{HAE in the Einstein Cross } \label{s4}

The Einstein Cross QSO 2237+0305 \citep{Huchra_85}
consists of a quadruply imaged quasar and a lensing galaxy that is
the nearest of all known gravitational lens systems ($z_G =
0.0395$). The gravitational delay times between images in this GLS
are  of the order of hours. This follows from a highly symmetric configuration of the images and is partially confirmed by observations (see, e.g., \citealt{Schmidt_98, Dai_03,
Vakulik_06, Fedorova_09}). Since  the Einstein Cross is a very
suitable object for microlensing studies, its  images  have been continuously
monitored by different groups for more than a dozen of years.
In this system, significant microlensing-induced brightness peaks on light curves
of the quasar images were detected  (see, e.g.,
\citealt{wozniak_00, alcald_02, Moreau_05}).

\subsection*{Fitting the light curves and estimations of HAE
parameters in GLS Q2237+0305}

\noindent We now apply formula~(\ref{eq18}) to the fitting
 of the light curves near  HAE.
For a moving source, $Y_{i} = V_i \left( {t - t_C } \right)$,
where $t$ is the time, $t_C $ is the time of the crossing of the caustic
by the source centre, $V_i $ is the projection of the source
velocity on the axis $y_i $. We suppose that  $V_2 >
> V_1 $, i.e. the source crosses the caustic effectively and does not
move along it. Our numerical simulations have shown that the terms
depending upon the coordinate  $s$ contribute only for small
angles between the source trajectory and the tangent to the
caustic. Therefore, we do not take them into
account, and, correspondingly, the parameter  $Q$ is not involved
into consideration. Introducing the parameter $T = L
\mathord{\left/ {\vphantom {L {\left| {V_2 } \right|}}} \right.
\kern-\nulldelimiterspace} {\left| {V_2 } \right|}$, we obtain $h = \pm
{\left( {t - t_C } \right)} \mathord{\left/ {\vphantom {{\left( {t
- t_C } \right)} T}} \right. \kern-\nulldelimiterspace} T$ (the
sign ``+'' corresponds to the source motion along the positive
direction of  the $y_2 $ axis).

We consider the known HAE in the light curve of image C of GLS Q2237+0305 using the
OGLE data recorded during 1999~\citep{wozniak_00}. Let $F_0 $ be the flux from image C when the
microlensing is absent. Under the supposition that the proper
brightness variations of the quasar in GLS can be neglected and taking  expression~(\ref{eq18}) for the amplification into account, we
obtain the formula for fitting the flux from the microlensed
Gaussian source,

\begin{equation}
\label{eq21}
F^M\left( t \right) = A + B\Phi _0 \left( h \right) + C\Phi _1\left( h
\right) + D\Phi _2 \left( h \right),
\end{equation}

\noindent which contains the parameters
\[
\begin{array}{l}
 A = F_0 K_0 , \quad
 B = {F_0 } \mathord{\left/ {\vphantom {{F_0 } {\sqrt {4\pi T\left| b
\right|\left| {V_2 } \right|} }}} \right.
\kern-\nulldelimiterspace} {\sqrt
{4\pi T\left| b \right|\left| {V_2 } \right|} }, \quad
 C = F_0 P\sqrt {{T\left| {V_2 } \right|} \mathord{\left/ {\vphantom
{{T\left| {V_2 } \right|} {4\pi \left| b \right|}}} \right.
\kern-\nulldelimiterspace} {4\pi \left| b \right|}} , \quad
 D = - \frac{1}{4}F_0 \kappa \sqrt {{T\left| {V_2 } \right|} \mathord{\left/
{\vphantom {{T\left| {V_2 } \right|} {\pi \left| b \right|}}}
\right. \kern-\nulldelimiterspace} {\pi \left| b \right|}}
 \end{array},
\]
\noindent and $t_C \mbox{ }$ and $T$ which
appear nonlinearly. The quantity $K_0 $ in the expression
for $A$ is a part of the amplification due to noncritical
images. The parameters $A$ and $B$ are evidently positive, $D$ is
negative, and $C$ can have values of both signs. As discussed above, the possibility to use the linear caustic approximation or formula (\ref{eq18})  is determined by
the ratios of corrections coefficients to the coefficient $B$ of
the zeroth approximation: $C \mathord{\left/ {\vphantom {C {B =
LP,}}} \right. \kern-\nulldelimiterspace} {B = LP,} \quad D
\mathord{\left/ {\vphantom {D {B = { - L\kappa } \mathord{\left/
{\vphantom {{ - L\kappa } 2}} \right. \kern-\nulldelimiterspace}
2}}} \right. \kern-\nulldelimiterspace} {B = { - L\kappa }
\mathord{\left/ {\vphantom {{ - L\kappa } 2}} \right.
\kern-\nulldelimiterspace} 2}$.

To fit the light curve, we used the minimization of the weighted sum of squares:

\begin{equation}
\label{eq22} S = \sum\limits_{i = 1}^N {W_i \left[ {F_i -
F^M\left( {t{ }_i} \right)} \right]^2} ,
\end{equation}

\noindent where $F_i$ is the result of the
$i$-th measurement, and $W_i = 1 \mathord{\left/ {\vphantom {1 {\sigma
_i^2 }}} \right. \kern-\nulldelimiterspace} {\sigma _i^2 }$ is its weight that is expressed through the corresponding
dispersion estimate $\sigma _i $  \citep{wozniak_00}. The fitting quality is often characterized by the
parameter $\chi ^2 = {S_{\min } } \mathord{\left/ {\vphantom
{{S_{\min } } \nu }} \right. \kern-\nulldelimiterspace} \nu $,  $\nu $ being the number of degrees of freedom. The value of this parameter
 in the optimal case should tend to 1.

As $F^{M}$, we have considered  the following
models:

\[
F^0\left( t \right)= A + B\Phi _0 \left( h \right),\vspace{2\jot}
\]
\[
F^1\left( t \right) = A + B\Phi _0 \left(
h \right) + C\Phi _1 \left( h \right), \vspace{2\jot}
\]
\[
F^2\left( t \right) = A + B\Phi _0 \left( h \right) + D\Phi _2
\left( h \right).
\]

\noindent  We also analysed the model that takes both
correction terms $C\Phi _1 \left( h \right)+D\Phi _2
\left( h \right)$ into account. However, we found that it does not allow us to obtain the
coefficients that are statistically significant simultaneously. For comparison, we also considered the model with a correction term linear in $h$; such a correction can be caused by the own variability of the quasar, or by the influence of noncritical images (cf. \citealt {yonehara_01}):

\[
F^3\left( t \right) = A + B\Phi _0
\left( h \right) + Eh.
\]

The results of best-fitting with different models are presented in Table~\ref{tabl}. It contains  the estimates of model parameters
and their central 95-per-cent  confidence intervals that
have been found by the Monte-Carlo simulations under supposition
of the normal distribution of errors.
In all the models, the correction terms are statistically
significant. The probability that the correction
coefficient is occasionally nonzero is certainly less than $10^{ - 3}$ in
every case. On the other hand, all three models can compete with one another on
an equal footing (and probably with another effects such as those due to
a complicated source structure).

\begin{table}

\caption{Model parameters and their central 0.95 confidence intervals as the results of light curve fitting for Q2237+0305C. \newline The data due to OGLE group  \citep {wozniak_00}. }
\label{tabl}
\begin{tabular}{c|c|c|c|c|c|c|c|c} \hline
{Fitting}&$t_C $ &$T$&$A$&$B$&{Correction}&$S_{min}$&$\nu$&$\chi^2$\\
 {model}&(JD-2450000)&{(days)}&{(mJy)}&{(mJy)}&{coefficient  (mJy)} & \\ \hline\hline
$F^0$   &${1384.7}_{ - 1.0}^{ + 1.0}$&$33.0_{ - 2.6}^{ + 2.7} $&
${0.364}_{ - 0.006}^{ + 0.006} $&
${0.076}_{ - 0.003}^{ + 0.003} $& & 53.91 & 40 & 1.35\\

$F^1$  &${1382.6}_{ - 1.7}^{ + 1.5} $& $34.7_{ - 2.7}^{ + 2.9} $&
${0.366}_{ - 0.006}^{ + 0.006} $& ${0.079}_{ - 0.004}^{ + 0.004}
$& $C = - 0.006_{ - 0.004}^{ + 0.003} $& 38.43
 & 39 & 0.99\\

$F^2$&${1367.1}_{ - 3.4}^{ + 4.4} $& $40.1_{ - 3.5}^{ + 3.2} $&
${0.363}_{ - 0.007}^{ + 0.006} $& ${0.067}_{ - 0.004}^{ + 0.005}
$& $D = - 0.071_{ - 0.010}^{ + 0.014} $& 34.97
 & 39 &0.90\\

$F^3$&${1383.3}_{- 1.4}^{ + 1.3} $& $34.9_{ - 2.6}^{ + 3.2} $&
${0.358}_{- 0.008}^{ + 0.007} $& ${0.080}_{- 0.004}^{ + 0.005} $&
$E = - 0.004_{ - 0.003}^{ + 0.002} $& 39.38 & 39 &1.01\\
\hline
\end{tabular}

\medskip
Notation for the fitting models and the parameters are given in the text.
\end{table}

Note that the flux variation of the model light curve is roughly equal to 0.18 mJy, and the standard deviation of data is $\sigma \approx 0.006$ mJy. Thus, the assumption of the 5-per-cent tolerance in a vicinity of the light curve maximum, which has been used in criteria~(\ref{LCA}), is  rather realistic. In the case of the $F^1$-model, we find $C/B=LP\approx0.076$.  The comparison with the first inequality in (\ref{LCA}) indicates the agreement with the statistical significance of the first correction term.

\begin{figure}
    \centering
        \includegraphics{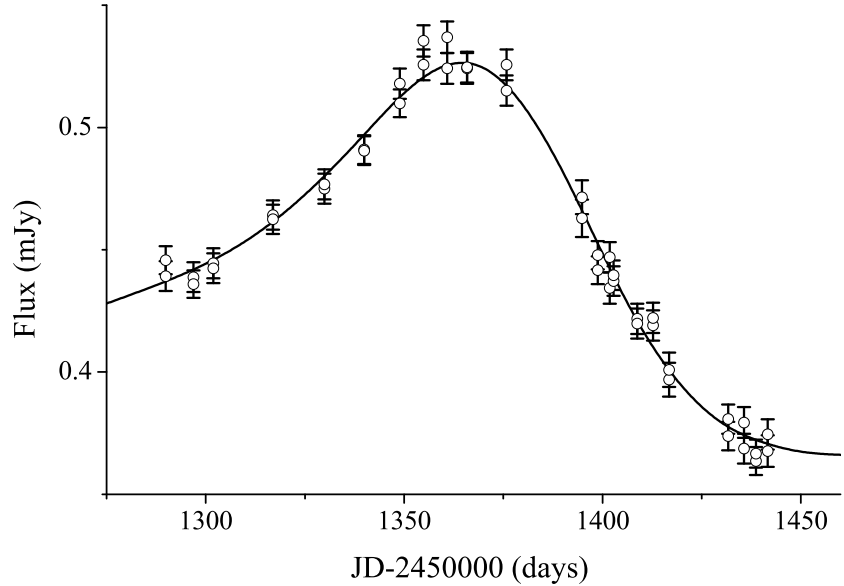}
    \caption{Observed light curve of Q2237+0305C (OGLE data) and the best-fitting with the model $F^1$ (solid line).}
    \label{Fi3}
\end{figure}

Fig.~\ref{Fi3} shows the results of light curve fitting  for image
C in the region of HAE with the $F^1$-model. Here, 44 data
points were included corresponding to the [1289,1442] epoch interval
(light circles).  To choose a fitting interval in a vicinity of the light curve maximum, we used the fact that the values of fitting parameters are reasonably stable with respect to  a reduction of the interval. This allowed
us to find a preliminary estimate of the ``background level'' $A$.
After that, all the points above $A$ have been involved in the final treatment.

The models  $F^1$ and $F^3$ fit the data equally well.
Let us discuss the model  $F^2$ and the role of the
second correction. As compared with the
other models,  the model $F^2$ leads to a somewhat lower value of $\chi ^2$,
but the confidence interval for $t_C \mbox{ }$ is four times wider
than that in the case of the model $F^0$, and the source size appears to be
20 per cent more than that within other models. Here, the typical features of the
second correction mentioned in the previous section become
apparent. The true determination of  $D$ from  formula~(\ref{eq21}) could allow us to estimate the ratio of the curvature and
source radii: $\kappa L = - {2D} \mathord{\left/ {\vphantom {{2D}
B}} \right. \kern-\nulldelimiterspace} B$. But the obtained value of $D$  corresponds to the curvature
radius  $R_{c}\equiv 1/\kappa$ that is about twice less than the
source ``size''  $L$. So, the condition of smallness of the correction is violated, and the model $F^2$ is not
acceptable. On the other hand, neglecting the second correction in (\ref{eq21})
(i.e., the transition to the model $F^1$) corresponds to $R_{c}>>L$. Our
calculations have shown that the \textsl{a priori} assignment $R_{c}>L$ and the introduction of the appropriate additional term into the model $F^1$ have little influence on the estimates of $A,B$, and $C$.

To sum up, the model  $F^1$ fits satisfactorily  the data
on HAE in question within the present accuracy, though we cannot
rule out competing models and/or competing effects.

We also analysed HAE in the light curve of image Q2237+0305A
that happened in 1999 \citep{wozniak_00}. However, the treatment
of this HAE does not allowed us to find statistically
significant estimates of the
corrections because of lacking the data corresponding to the
 caustic inner region.

\section{Discussion}
\noindent We proposed two methods that enabled us to obtain the
critical solutions of the gravitational lens equation near a fold
caustic with any desired accuracy. The first method is based on the expansion of the solutions and the
 amplification factor $K_{cr} $ of two critical images of a point source  in powers of
 a formal parameter $t$ that can be interpreted as a parameter describing the proximity to the
 caustic. We derived a representation of $K_{cr} $ which is convenient
 to obtain the amplification factor for an extended source.
This representation is not analytic in local coordinates, and  $K_{cr} $
contains a nonintegrable term. In order to elucidate this term, we considered a different representation of the solutions which contains analytic functions of coordinates and the parameter $t$, as well as square roots of such functions. This method yields only integrable
expressions and explains the meaning of the nonintegrable term in $K_{cr} $: it should be
considered as a distribution which can be used to
calculate the amplification of a small (and smooth) extended source near
the fold caustic by taking the appropriate convolution.

In order to obtain nontrivial corrections to  $K_{cr}$, we had to
use higher orders of the expansion of the lens equation as compared to
works by \cite{alzh_03} and  \cite{keeton_05}. This is a
consequence of the cancellation of terms $\sim O\left( t^2
\right)$ which are present in the amplifications of separate critical
images. Therefore, we had to deal with terms of the order $\sim
O\left( {t^3} \right)$ and the Taylor expansion including the 4-th
order in the lens equation (\ref{eq3}). The modified formula for
$K_{cr} $ contains 3 extra parameters that are combinations of 5
coefficients of the Taylor expansion (we confined ourselves to the
case of no continuous matter on the line of sight).

Based on our result concerning $K_{cr}$, we derived an asymptotic formula
for the amplification $K_G$ of a small Gaussian source. We analysed
the role of  new
 ``post-linear'' corrections and formulated the conditions of applicability of the linear caustic approximation. The fitting
 of the light curve of GLS Q2237+0305C  shows
 that some of these corrections can be statistically significant.

Of course, we are far from the thought that the available observational
data make it possible to determine all additional
parameters  simultaneously (and even more so, the coefficients of the expansion of the
lens mapping up to the orders involved). Such a determination
requires a considerable improvement of the photometric accuracy in the
future. Our result is more modest: we have shown that, even at the present
level of accuracy, some elements of the light curves require to consider
the higher-order terms in the solutions of the lens equation. At least, after the introduction of some
post-linear corrections, $\chi^2$ decreases by 30 per cent  thus approaching 1. One may hope that an enhancement of the observational accuracy  will
increase the role of post-linear approximations.

In this connection, we note that, besides estimating the source
size  in GLS on the basis of light curves, a number of
authors \citep{shalyapin_01,shalyapin_02,Bogdanov_02, goico_01,Kochanek_04,
Mortonson_05,gil_06, vakulik_07,anguita_08} discussed some delicate questions concerning a fine
quasar structure. For example, \cite{goico_01} wrote that the
GLITP data \citep{alcald_02} on GLS Q2237+0305 admit only
accretion disc models (see also \citealt{gil_06, anguita_08}).
Obviously, the presence of an accretion disk in a central region of
quasar is beyond any doubts, as well as the fact that the real
appearance of the quasar core can be quite different from
very simplified theoretical models in question. On the other hand, \cite{Mortonson_05} argue
that the accretion disk can be modelled with any brightness
profile (Gaussian, uniform, etc.), and this model will agree with the
available data provided that an appropriate source size is chosen.
Anyway, in these considerations, different effects  come into
play, and the consistent treatment requires to consider
all the corrections to the amplification including those
obtained in the present paper.

\section*{Acknowledgments}

This work has been supported in part by the ``Cosmomicrophysics"
programme of the National Academy of Sciences of Ukraine. VIZ also
acknowledges the partial support of Swiss National Science Foundation (SCOPES grant 128040). 
We would also like to thank anonymous referees for stimulating comments that helped us to improve the text and initiated the writing of Appendices \ref{C}
and \ref{D}.

\appendix
\section[A]{Analytical expansion method}
\label{A}

Here, we present a justification of the method of analytical
expansions near the folds used in Section \ref{ss2.1} following
\cite{alzh_03}. This problem is not trivial, because the Jacobian
of the lens mapping on the fold is equal to zero.

We say that an analytic function $f(t)$ has order $k$ at $t=0$ , if $f^{(i)}(0)=0$ for $i=0,1,...,k-1$, and $f^{(k)}(0)\neq0$ (\citealt{Poston_78}); then we write $f(t)= O(t^k)$.

It is well known that the mapping

 \begin{equation}
\label{eq0a}
{\bmath y} = {\bmath F}({\bmath x})
\end{equation}

 \noindent of two-dimensional manifolds (e.g.,  mapping~(\ref{eq3})) in a neighborhood
of the fold ($\bmath x=\bmath y=0$) can be reduced by the coordinate transformations
\begin{equation}
\label{eq2a} {\rm {\bmath u}} = {\rm {\bmath u}}\left( {\rm
{\bmath x}} \right), \quad {\rm {\bmath v}} = {\rm {\bmath
v}}\left( {\rm {\bmath y}} \right),
 \quad {\rm {\bmath x}} = {\rm
{\bmath x}}\left( {\rm {\bmath u}} \right), \quad {\rm {\bmath y}}
= {\rm {\bmath y}}\left( {\rm {\bmath v}} \right)
\end{equation}
\noindent
to the normal form (see, e.g., \citealt
{Poston_78,Petters}):

\begin{equation}
\label{eq1a}
{\rm {\bmath u}} \to {\bmath v} :\quad v_1 = u_1 \mbox{ , }v_2 = \left( {u_2 }
\right)^{2}.
\end{equation}

\noindent  We
suppose the initial mapping~(\ref{eq0a}) is analytic;
then transformations~(\ref{eq2a}) are also analytic.

Let the source move along a parameterized curve ${\rm {\bmath v}}\left( t
\right)$ such that (i) $v_2 \left( t \right) > 0$ for $t \ne 0$,
${\rm {\bmath v}}\left( 0 \right) = 0$; (ii) the functions $v{ }_i\left( t
\right)$ are analytic at $t = 0$; (iii) the order of  $v_2 \left( t
\right)$ is $2k$ for some integer $k \geq 1$. Conditions (ii) and (iii) mean that $v_2 \left( t \right) = t^{2k}\left(
{q + \chi \left( t \right)} \right), \mbox{ }q \ne 0$, and $\chi \left( t
\right)$ is analytic function at $t = 0$ so that $\chi \left( 0 \right) =
0$. In view of (i), we have $q + \chi \left( t \right) > \mbox{0}$.

We have two obvious solutions of Eqs.~(\ref{eq1a}):

\begin{equation}
\label{eq5a} \bmath u_{( \pm )}(t)=\{v_1(t),\pm \, t^{k}\sqrt
{q + \chi ( t )}\},
\end{equation}

\noindent
 which, in view of $q>0$, are analytic as functions of $t$ at the common point $t=0$. We can say that the trajectory of the source consists of two branches, corresponding to values $t<0$ and $t>0$. Each branch has two images lying on different sides of the critical curve. When $k$ is even (odd), the images which lie on the one side (on different sides) continue analytically  one another.
 Thus, a suitable choice of the source trajectory  $\bmath v(t)$ leads to the analyticity of its critical images as functions of the parameter.

We  now intend to reformulate the sufficient conditions of the existence of analytical solutions in the original coordinates.
The coordinate transformation $\rm {\bmath y}\rightarrow\rm
{\bmath v}$  in the source plane can be written as

\begin{equation}
\label{eq6aa}
 v_1 = Ay_1 + By_2 + \Phi _1 \left( {y_1 ,y_2 } \right),
\end{equation}

\begin{equation}
\label{eq6a}
 v_2 = Cy_1 + Dy_2 + \Phi _2 \left( {y_1 ,y_2 } \right),
\end{equation}
\[
 AD - BC \ne 0.
\]

Here, $\Phi _i \left( {y_1 ,y_2 } \right)$ are analytic functions
having the Taylor expansions starting from the second order.

Then we assume that the terms linear in $x_i $  are present only in the
first component, $F_1( \bmath x)$, of the lens mapping~(\ref{eq0a}).
Then, in Eq.~(\ref{eq6a}),  $C = 0$.
 This statement follows from the substitution of (\ref{eq0a}) in
 (\ref{eq6a}), the application of the transformation ${\rm {\bmath x}} = {\rm
{\bmath x}}\left( {\rm {\bmath u}} \right)$, and the comparison with the second equation in (\ref{eq1a}).

\textit{Suppose that}

\textit{(a) the curve $\bmath y(t)$ lies on the inner side
of the caustic (except $\bmath y(0)=0$), }

\textit {(b) $\bmath y(t)$ is analytic and such that $y_1(t)=O(t^m),       y_2(t)=O(t^{2k})$ ; $m>k\ge 1$, and $m$ and $k$ are integers.}

\textit {Then Eq.~(\ref{eq6a}) yields $v_2(t)=O(t^{2k})$ and conditions (i),(ii), and (iii) are fulfilled.}

In summary, we obtain the following statement.

\textit{Let the function $\bmath F( \bmath x)$ of the lens
equation~(\ref{eq0a}) be an analytic function of $\bmath x$  in
a neighborhood of the fold critical point $\bmath x = 0$, let
$\bmath y = 0$ be the corresponding caustic point, and let $F_2(\bmath x)$ do not contain linear terms. Let also $\bmath y(t)$ be an analytic vector-function which represents the curve satisfying conditions (a),(b)}.

\textit{Then the lens equation~\ref{eq0a} has two  solutions ${\bmath x}_{(+)}( t )$ and  ${\bmath x}_{(-)}( t )$ which are analytic in $t$ and
represent the critical images of the curve $\bmath y(t)$ }.

These conditions are sufficient ones, and they do not exhaust possible combinations of $k,m$. Typically, the statement is true also when $m = k$.
This statement allows us to use analytical expansions along
test
 source trajectories to obtain the critical solutions near the
folds.
In the
main body of this paper, we have used the family of straight-line
trajectories of the source $y_i ( t ) = a_i t^2$ ($k = 1, m=2$) in a neighborhood of the
caustic on one side from the tangent to the caustic.
In work \citep{alzh_03}, we also considered the version with
parabolic curves $y_1 \left( t \right) = at$, $y_2 \left( t
\right) = bt^2,$ which allowed us to find the solutions in the
approximation of a parabolic caustic.

\section[B]{Analytical iteration method }
\label{B}

 In this Appendix, we show that the critical solutions of the lens
 equation~(\ref{eq3}) (after the substitution of (\ref{eq4})) can be represented in the form~(\ref{new1}), the functions $p,r,\bar s,w$ being polynomials
 in $t,y_1$, and $y_2$ on every step of the
 approximation procedure.

Here, we use the  notations which are independent of the rest of the paper. After a simple change of the variables ($y_1=(b\y_1-a\y_2)/2b$, $y_2=-\y_2/b$), Eq.~(\ref{eq25}) can be written in the form

\[
 y_1 = x_1 + t\sum\limits_{n,m} {a_{n,m}(t) x_1^n x_2^m } \;,
 \]

\begin {equation}
\label{eq1b}
 y_2 = x_2^2 + t\sum\limits_{n,m} {b_{n,m}(t) x_1^n x_2^m } \;,
\end {equation}

\noindent where the indices $m,n$ take on integer nonnegative
values; $t$ is a small parameter; the coefficients $a_{n,m}(t)$  and $b_{n,m}(t)$ are finite-order polynomials in $t$. In fact, all
considerations below can be performed in the case where the r.h.s. of
(\ref{eq1b}) is an analytic functions in $x_1 ,x_2 $, and the  small
parameter $t$. However, in this paper, we deal with finite-order
approximations.

We search for a solution in the form

\begin {equation}
\label{eq2b}
x_1 = p + t\varepsilon r\sqrt w ,\quad x_2 = ts
+ \varepsilon \sqrt w ;\quad \varepsilon = \pm 1.
\end {equation}

\noindent After the substitution of Eq.~(\ref{eq2b}) in (\ref{eq1b}), we
separate the terms containing integer and half-integer powers of $w$, e.g.,

\[
 \sum\limits_{n,m} {a_{n,m} } (p + \varepsilon tr\sqrt w )^n(ts +
\varepsilon \sqrt w )^m = A_0+\varepsilon w^{1/2} A_1,
\]
\*where

\*$A_0=\sum\limits_{n,m} {a_{n,m} }
\sum\limits_{\scriptstyle {k,k'}\atop \scriptstyle{k + k' = 2K} } {C_n^k C_m^{k'} } (tr)^k
p^{n - k}(ts)^{m - k'}w^{K},$ \qquad  $K=\frac{k+k'}{2}$ \  is an integer,

\begin {equation}
\label{eq31b}
A_1=\sum\limits_{n,m} {a_{n,m} }
\sum\limits_{ \scriptstyle {k,k'}\atop  \scriptstyle{k + k' = 2\bar{K}+1 }} {C_n^k C_m^{k'} } (tr)^k
p^{n - k}(ts)^{m - k'}w^{\bar{K}}$,
\* \qquad  $\bar{K}=\frac{k+k'-1}{2}$ \ is an integer.
$\end {equation}

\noindent Here, $C_n^k $ are the binomial coefficients, $C_n^k = 0$
for $k > n$ and for $k<0$; and $k,k'$ are nonnegative integers.

Then we equate separately the terms with integer and half-integer
powers of $w$ on both sides of the relation following from the
first equation of system~(\ref{eq1b}) and obtain

\begin {equation}
\label{eq3b}
p = y_1 + tP(t,p,r,s,w),
\end {equation}

\noindent where

\[
P(t,p,r,s,w) \equiv - A_0,
\]

\*and

\begin {equation}
\label{eq-for-r}
r + A_1=0.
\end {equation}

Analogously, the second equation of (\ref{eq1b}) yields

\begin {equation}
\label{eq5b}
w = y_2 + t W(t,p,r,s,w),
\end {equation}

\noindent where
\[
W(t,p,r,s,w) \equiv - ts^2 - B_0,
\]

\noindent and

\begin {equation}
\label{eq-for-s}
2s+B_1=0.
\end {equation}

\*Here, $B_0$ and $B_1$ are defined  similarly to $A_0$ and $A_1$ by  Eq.~(\ref{eq31b}) with the replacement of the coefficients, $a_{n,m}\rightarrow b_{n,m}$.

Separating  the terms of the zero order with respect to $r$ and $s$ (or with respect to $t$) in Eq.~(\ref{eq31b}), we rewrite Eq.~(\ref{eq-for-r}) as follows:

\begin {equation}
\label{eq-for-r1}
r + \sum\limits_{n,K} {a_{n,2K+1} } p^n w^K +
tU(t,p,r,s,w)=0.
\end {equation}

\*Here,

\[
U(t,p,r,s,w)\equiv\sum\limits_{n,m} {a_{n,m} } \sum \limits_{ \scriptstyle{k,k'}\atop  {\scriptstyle{k +
k'=2K+1}\atop \scriptstyle {k+m-k'>0}}} C_n^k C_m^{k'} t^{k + m - k' - 1} r^k p^{n - k} s^{m -
 k'}w^{K}.
\]

Substituting $p$ and $w$ from Eqs.~(\ref{eq3b}) and (\ref{eq5b}) in
Eq.~(\ref{eq-for-r1}), we get

\[
 r + \sum\limits_{n,K} {a_{n,2K + 1} } (y_1 + tP)^n(y_2 + tW)^{K}
 + tU(t,p,r,s,w)=0.
\]

This relation can be written as

\begin {equation}
\label{eq7b}
r = f(t,y_1 ,y_2 ) + t R(t,p,r,s,w),
\end {equation}

\noindent where

\[
f(t,y_1 ,y_2 ) \equiv - \sum\limits_{n,K} {a_{n,2K + 1} y_1 ^n y_2^K},
\]

\[
 R(t,p,r,s,w) \equiv - \sum\limits_{n,K} {a_{n,2K + 1} }
\sum\limits_{ \scriptstyle{k,k'}\atop
 \scriptstyle{k+k' >0}} {C_n^k C_{K}^{k'} } y_1 ^{n - k}y_2 ^{K - k'}P^kW^{k'}t^{k +
k' - 1} - U(t,p,r,s,w).
\]

An analogous consideration of  Eq.~(\ref{eq-for-s}) yields

\[
s + \frac{1}{2}\sum\limits_{n,K} {b_{n,2K + 1} } p^nw^K +
\frac{t}{2}V(t,p,r,s,w)=0,
\]

\*where

\[
V(t,p,r,s,w)\equiv\sum\limits_{n,m} {b_{n,m} } \sum \limits_{ \scriptstyle{k,k'}\atop  {\scriptstyle{k +
k'=2K+1}\atop \scriptstyle {k+m-k'>0}}} C_n^k C_m^{k'} t^{k + m - k' - 1} r^k p^{n - k} s^{m -
 k'}w^{K}.
\]

Then we substitute $p$ and $w$ from Eqs.~(\ref{eq3b}) and (\ref{eq5b}) to obtain

\begin {equation}
\label{eq8b}
s = g(t,y_1 ,y_2 ) + t S(t,p,r,s,w),
\end {equation}

\noindent where

\[
g(t,y_1 ,y_2 ) \equiv - \displaystyle {\frac{1}{2}} \sum\limits_{n,K} {b_{n,2K + 1} } y_1^n y_2^K,
\]
\[
S(t,p,r,s,w) \equiv - \displaystyle {\frac{1}{2}}\left[\sum\limits_{n,K} {b_{n,2K + 1} }
\sum\limits_{ \scriptstyle{k,k'}\atop
 \scriptstyle{k+k' >0}} {C_n^k C_{K}^{k'} } y_1 ^{n - k}y_2 ^{K - k'}P^kW^{k'}t^{k +
k' - 1} + V(t,p,r,s,w)\right].
\]

Thus, we have the system of equations
(\ref{eq3b}), (\ref{eq7b}), (\ref{eq8b}), (\ref{eq5b}) for
$p,r,s$, and $w$ that can be represented in the form

\begin {equation}
\label{eq9b}
{\rm {\bmath X}} = {\rm {\bmath F}}(t,{\rm {\bmath Y}}) +
t\,{\rm {\bmath G}}({t,\rm {\bmath X}},{\rm {\bmath Y}}),
\end {equation}

\noindent where

\[
{\rm {\bmath X}} = \left\{ {\begin{array}{l}
 p \\
 r \\
 s \\
 w \\
 \end{array}} \right\},\;{\rm {\bmath Y}} = \left\{ {\begin{array}{l}
 y_1 \\
 y_2 \\
 \end{array}} \right\},\;{\rm {\bmath F}}(t,{\rm {\bmath Y}}) = \left\{
{\begin{array}{l}
 y_1 \\
 f(t,y_1 ,y_2 ) \\
 g(t,y_1 ,y_2 ) \\
 y_2 \\
 \end{array}} \right\},\;{\rm {\bmath G}}(t,{\rm {\bmath X}},{\rm {\bmath Y}}) =
\left\{ {\begin{array}{l}
 P(t,p,r,s,w) \\
 R(t,p,r,s,w) \\
 S(t,p,r,s,w) \\
 W(t,p,r,s,w) \\
 \end{array}} \right\}.
\]

System~(\ref{eq9b}) is ready for iterations

\[
{\rm {\bmath X}}_{(n)} = {\rm {\bmath F}}(t,{\rm {\bmath Y}}) + t\,{\rm
{\bmath G}}(t,{\rm {\bmath X}}_{(n - 1)} ,{\rm {\bmath Y}}),\quad n =
1,2,...,\quad {\rm {\bmath X}}_{(0)} = {\rm {\bmath F}}(t,{\rm {\bmath
Y}}).
\]

For a sufficiently small $t$, the iteration process converges due
to the contraction mapping theorem. It is worth to note that,
at every iteration step, we obtain an approximate solution in the
form of finite-order polynomials in ${\rm {\bmath Y}}$. This is
obvious from the explicit form of the functions $ f(t,y_1 ,y_2
), g(t,y_1 ,y_2 ), P(t,p,r,s,w), R(t,p,r,s,w), S(t,p,r,s,w)$,  and $W(t,p,r,s,w)$.
The application of the above procedure  to system (\ref{eq25}) yields
solution~(\ref{new1}-\ref{new1Z}).

\section[C]{amplification for a power-law source}
\label{C}

Here, we use Eq.~(\ref{flux_extended}) to derive the
magnification of an extended centrally symmetric source with a
power-law brightness distribution. Two different types of
``power-law" distributions can be found in the literature on the
gravitational lensing. In particular, there are  the distributions
 \citep{shalyapin_01,shalyapin_02}
\begin{equation} \label{C1}
 I^{(-)}_p ( {\mathbf y} ) = \frac{p - 1}{\pi L^2}\left[ 1 +
 {\mathbf y}^2 /L^2 \right]^{ - p},
\end{equation}
\*where $p > 1$ is the power index, the
source centre is at the coordinate origin, the distribution (\ref{C1}) is normalized to unity, and $L$ is related to the r.m.s. radius $R_{rms}$ as
$L^2 = \left( {p - 2} \right)R_{rms}^2 $,
\[
R_{rms} ^2 =  \int\ {dy_1dy_2 \, {\mathbf y}^2  I^{(-)}_p (
{\mathbf y} ) }.
\]
\* For fixed $R_{rms} $ and $p \to \infty $, the brightness
distribution (\ref{C1}) tends to the Gaussian one.

Along with  (\ref{C1}), the  models for limb darkening are also
often used in microlensing studies  (see, e.g., \citealt{dominik})
\begin{equation}
 \label{C2}
I_q^{( + )} ( {\bf y}) = \frac{q + 1}{\pi L^2} \, \Xi ( {|\mathbf
y|}/L;q ) , \quad \Xi ( \xi;q ) = \Theta (1 - \xi ^2)(1 - \xi
^2)^q ,
\end{equation}
\* where $L$ stands for the source radius, and $L^2 = \left( {q + 2}
\right)R_{rms}^2 $. Here, we assume $q > 0$. Linear combinations of
distributions (\ref{gaussian distribution}), (\ref{C1}), and (\ref{C2})
with different parameters yield rather a wide class of symmetric
source models.

For brightness profile (\ref{C1}),
the total microlensed flux (\ref{flux_extended}) that describes a
variable contribution of the critical images, as the source crosses
a fold caustic, is the convolution of (\ref{C1}) with
(\ref{generalized Kcr}). The result for the amplification factor
 involves integrals that can be expressed via the hypergeometric
function $_2F_1 $ \citep{bateman}:
\[
\Psi _{k,p} \left( h \right) = \frac{\Gamma \left( {p -
\frac{1}{2}} \right)}{\Gamma \left( {p - 1} \right)}
\int\limits_0^\infty {\frac{y^{k - \frac{1}{2}}dy}{\left( {1 +
\left( {y - h} \right)^2} \right)^{p - 1 / 2}}}=
\]
\begin{equation}
\label {powercoeff_res}
 = \frac{\Gamma
\left( {p - \frac{1}{2}} \right)}{\Gamma \left( {p - 1} \right)}
B\left( {k + \frac{1}{2},2p - k - \frac{3}{2}} \right)\, (1 +
h^2)^{k/2+3/4-p}\,{ }_2F_1 \left( {k + \frac{1}{2},2p - k -
\frac{3}{2};\,p\,\,;\frac{1}{2}\left( {1 + \frac{h}{\sqrt {1 +
h^2} }} \right)} \right)
\end{equation}

\*for $k = 0,1$, $B\left( {x,y} \right)$ being the Beta-function.

We extend (\ref{powercoeff_res}) to $k = -1$ having in mind the
definition of $(y)_+^{ - 3/2}$, so that

\begin{equation}
\label{eq3app_C} \Psi _{-1,p} ( h) = 4(p-1)[ h\Psi _{0,p+1} ( h) -
\Psi _{1,p+1} ( h)].
\end{equation}

\*Like in Subsection \ref{ss3.3}, we introduce the normalized
coordinates of the source center $ s = Y_{1} / L, \quad h =
Y_{2} / L$. Now, the amplification due to critical images takes on the form

\begin{equation}
\label{eq4app_C} K^-_p ( s,h ) =  \frac{1}{2\sqrt {\pi |b|L }
}\left\{ {\Psi _{0,p} ( h) + L\left[ {P \Psi _{1,p}(h) -
\frac{\kappa }{8(p-2) }\Psi _{-1,p-1}(h) + Q \,s\, \Psi _{0,p}(h)
- \frac{\kappa}{4}\, s^2\Psi _{-1,p}(h) } \right]} \right\}.
\end{equation}

\*The zeroth approximation to this formula has been derived by
\citet{shalyapin_01} .

In the case of the model  with limb darkening (\ref{C2}), the critical
images disappear when the source lies on the outer side of the
caustic (i.e., for $h < - 1$). The substitution of (\ref{C2}) and
(\ref{generalized Kcr}) in (\ref{flux_extended}) yields the
total amplification of critical images as

\begin{equation}
\label{K_plus}
 K_q^{(+)} ( {s,h} ) =
\frac{1}{2\sqrt {\pi | b|L} }\left\{ {{\rm X}_{0,q} \left( h
\right) + L\left[ {P{\rm X}_{1,q} \left( h \right) - \frac{\kappa
}{8\left( {q + 2} \right)}{\rm X}_{ - 1,q + 1} \left( h \right) +
Q\,s\,{\rm X}_{0,q} \left( h \right) - \frac{\kappa }{4}\,s^2{\rm
X}_{ - 1,q} \left( h \right)} \right]} \right\},
\end{equation}
\noindent where we denote
\[
{\rm X}_{k,q} \left( h \right) = \frac{\Gamma \left( {q + 2}
\right)}{\Gamma \left( {q + \frac{3}{2}}
\right)}\int\limits_0^\infty {y^{k - \frac{1}{2}}\,\,\Xi( y -
h;q+1/2)\,dy},\quad k=1,2.
\]
Like the previous analogous cases, we define
\[{\rm X}_{ - 1,q}
\left( h \right) = 4\left( {q + 1} \right)\left( {h{\rm X}_{0,q -
1} - {\rm X}_{1,q - 1} } \right).
\]
We have
\[
{\rm X}_{k,q} \left( h \right) = 2^{q + \frac{1}{2}}\left( {1 + h}
\right)^{q + k + 1}\frac{\Gamma \left( {q + 2} \right)\Gamma
\left( {k + \frac{1}{2}} \right)}{\Gamma \left( {q + k + 2}
\right)}{ }_2F_1 \left( { - q - \frac{1}{2},q + \frac{3}{2};q + k
+ 2;\frac{1 + h}{2}} \right),
\]
for $ - 1 < h < 1$ and
\[
{\rm X}_{k,q} \left( h \right) = \sqrt \pi \left( {h + 1}
\right)^{k - \frac{1}{2}}{ }_2F_1 \left( {q +
\frac{3}{2},\frac{1}{2} - k;2q + 3;\frac{2}{h + 1}} \right),
\]
\*for $h > 1$.

The functions  $\Psi _{0,p} $, $\Psi _{1,p} $, $\Psi _{ - 1,p} $,
and ${\rm X}_{0,q} $, ${\rm X}_{1,q} $, ${\rm X}_{ - 1,q} $ are
analogous to  $\Phi _0 $, $\Phi _1 $, $4\Phi _2 $ from (25),
respectively; they have a similar qualitative behavior and the same
asymptotics. Note that all formulas for the extended source models are
transformed into that for point-source amplification (\ref{generalized Kcr}),
as the source size tends to zero.
 
\section[D]{Explicit comparison of the first-order formulas with 
counterparts from the paper by Keeton, Gaudi \& Petters  (2005) }
\label{D} 
Here we compare our expressions for the first order corrections with that of Appendix A2 from \citep{keeton_05}, further KGP. This is especially relevant 
because KGP considers the general case of the lens equation without 
supposition on harmonic potential. 

Initial equations (A7-A8) of the lens mapping of KGP are as follows

\[
 \xi u_1 = K\theta _1 - \left( {3e\theta _1^2 + 2f\theta _1 \theta _2 + 
g\theta _2^2 } \right) - \left( {4k\theta _1^3 + 3m\theta _1^2 \theta _2 + 
2n\theta _1 \theta _2^2 + p\theta _2^3 } \right),
\] 

\[
 \xi u_2 = - \left( {f\theta _1^2 + 2g\theta _1 \theta _2 + 3h\theta _2^2 } 
\right) - \left( {m\theta _1^3 + 2n\theta _1^2 \theta _2 + 3p\theta _1 
\theta _2^2 + 4r\theta _2^3 } \right).
\]
These equations need to be compared to our Eqs. (\ref{eq3}-\ref{eq25}).  
The correspondence between the coordinate notations is $\theta _i 
\leftrightarrow x_i ,u_i \leftrightarrow \tilde {y}_i ,\xi 
\leftrightarrow t^2$. 

In the general case the lens potential $\Phi \left( {\bmath x} \right)$ obeys the 
equation 

\begin{equation}
\Delta \Phi = 2\sigma \left( {\bmath x} \right),
\end{equation}

\noindent
where $\sigma \left( {\bmath x} \right)$ is the normalized surface density of the continuous matter. Coordinates are chosen to diagonalize the symmetric matrix $A_{ij}=\left.\partial{y_i}/\partial{x_j}\right|_{{\bmath x}=0}$. Its eigenvalues are $\lambda_1=1-\sigma(0)+\Gamma(0)$ and $\lambda_2=1-\sigma(0)-\Gamma(0)$, where $\Gamma$ is modulus of the complex shear  (see, e.g., \citealt{schneider_92}). At the critical point one of the eigenvalues is zero (the coordinates are chosen so that $\left.\partial{y_2}/\partial{x_2}\right|_{{\bmath x}=0}=0$), in this case we have   
 $\left.\partial{y_1}/\partial{x_1}\right|_{{\bmath x}=0}=K=2\left[ {1 - \sigma 
\left( 0 \right)} \right]$. 

Now we 
consider the correspondence of the expansion coefficients. For example Eq. 
(\ref{eq1}) yields $3e = \frac{1}{2}\Phi ,_{111} $; $g = \frac{1}{2}\Phi ,_{122} $. 
Under the assumption that during HAE $\sigma = \sigma _0 = const$ we have 
$3e=-g\;\to\;-a$. Analogously $f=-3h\;\to\;-b$, $4k=-\frac{2}{3}n=4r\;\to\;-c$, $p=-m\;\to\;d$.

For critical solutions near a fold caustic, KGP found Eqs. (A14-A15) which 
we repeat below along with our analogs

\[
 \theta _1^\pm = \frac{3hu_1 - gu_2 }{3hK}\xi + O\left( \xi \right)^{3 
\mathord{\left/ {\vphantom {3 2}} \right. \kern-\nulldelimiterspace} 
2}\mbox{ } \to \mbox{ }x_1 = \tilde {x}_1 t^2 = \frac{b\tilde {y}_1 - 
a\tilde {y}_2 }{2b\left( {1 - \sigma _0 } \right)}t^2 + O\left( {t^3} 
\right), 
\]
\begin {equation}
 \theta _2^\pm = \mp \sqrt {\frac{ - u_2 }{3h}} \xi ^{1 \mathord{\left/ 
{\vphantom {1 2}} \right. \kern-\nulldelimiterspace} 2} - \frac{3ghu_1 - 
\left( {g^2 + 2Kr} \right)u_2 }{9h^2K}\xi + O\left( \xi \right)^{3 
\mathord{\left/ {\vphantom {3 2}} \right. \kern-\nulldelimiterspace} 
2}\mbox{ } \to 
\end {equation}
\[
 x_2 = \tilde {x}_2 t = - \varepsilon \sqrt {\frac{ - \tilde {y}_2 }{b}} t - 
\frac{ab\tilde {y}_1 - \left[ {a^2 - c\left( {1 - \sigma _0 } \right)} 
\right]\tilde {y}_2 }{2b^2\left( {1 - \sigma _0 } \right)}t^2 + O\left( 
{t^3} \right).  
 \]

For $\sigma _0 = 0$ these expressions are equivalent to $\tilde {x}_1 = 
\tilde {x}_{10} + O\left( t \right)$, $\tilde {x}_2 = \tilde {x}_{20} + 
\tilde {x}_{21} t + O\left( {t^2} \right)$, where $\tilde {x}_{10} ,\tilde 
{x}_{20} ,\tilde {x}_{21} $ are given by (\ref{eq6}-\ref{eq5x}). Thus (from the 
viewpoint of our approach) KGP derives the first coordinate in zero 
approximation. However, as KGP point out this is sufficient to find the 
first order correction for the amplification. Namely, for the corresponding 
Jacobian of the lens mapping, KGP found (A17):

\[
 \left( {\mu ^\pm } \right)^{ - 1} = \pm 2K\sqrt { - 3hu{ }_2} \xi ^{1 
\mathord{\left/ {\vphantom {1 2}} \right. \kern-\nulldelimiterspace} 2} + 
\frac{4}{3h}\left( {g^2 - 3fh + 2Kr} \right)u_2 \xi + O\left( \xi \right)^{3 
\mathord{\left/ {\vphantom {3 2}} \right. \kern-\nulldelimiterspace} 
2}\mbox{ } \to 
\]
\begin {equation} 
 J = 4\varepsilon t\left( {1 - \sigma _0 } \right)\sqrt { - b\tilde {y}_2 } 
+ \frac{4}{b}t^2\left[ {a^2 + b^2 - c\left( {1 - \sigma _0 } \right)} 
\right]\tilde {y}_2 + O\left( {t^3} \right). 
\end {equation} 

For $\sigma _0 = 0$ the latter equation is the same (in the first 
approximation) as Eq.(\ref{eq10}). 

For $\sigma _0 \ne 0$ the approximate solutions follow from the 
corresponding ones with $\sigma _0 = 0$ by means of the rescaling 
\begin{equation}
\tilde {y}_i \to {\tilde {y}_i } \mathord{\left/ {\vphantom {{\tilde {y}_i } 
{\left( {1 - \sigma _0 } \right)}}} \right. \kern-\nulldelimiterspace} 
{\left( {1 - \sigma _0 } \right)},\mbox{ }a \to a \mathord{\left/ {\vphantom 
{a {\left( {1 - \sigma _0 } \right)}}} \right. \kern-\nulldelimiterspace} 
{\left( {1 - \sigma _0 } \right)},\mbox{ }b \to b \mathord{\left/ {\vphantom 
{b {\left( {1 - \sigma _0 } \right)}}} \right. \kern-\nulldelimiterspace} 
{\left( {1 - \sigma _0 } \right)},\mbox{ }c \to c \mathord{\left/ {\vphantom 
{c {\left( {1 - \sigma _0 } \right)}}} \right. \kern-\nulldelimiterspace} 
{\left( {1 - \sigma _0 } \right)},\mbox{ }...
\end{equation}
Analogous rescaling for the Jacobian is $J \to J\left( {1 - \sigma _0 } 
\right)^2$; it is this rescaling of the variables has been mentioned in 
Section \ref{s1}.

\label{lastpage}

\end{document}